\documentclass[nohyper,12pt,letterpaper]{JHEP3}
\usepackage[dvips]{epsfig}
\usepackage{amsfonts,amssymb}
\newcommand{\be}{\begin{equation}}
\newcommand{\ee}{\end{equation}}
\newcommand{\ben}{\begin{displaymath}}
\newcommand{\een}{\end{displaymath}}
\newcommand{\bea}{\begin{eqnarray}}
\newcommand{\eea}{\end{eqnarray}}
\newcommand{\bean}{\begin{eqnarray*}}
\newcommand{\eean}{\end{eqnarray*}}

\newcommand{\ads}[1]{\mbox{${AdS}_{#1}$}}
\newcommand{\adss}[2]{\mbox{$AdS_{#1}\times {S}^{#2}$}}

\newcommand{\commentout}[1]{}

\newcommand{\beq}{\begin{equation}}
\newcommand{\eeq}{\end{equation}}
\newcommand{\beqr}{\begin{displaymath}}
\newcommand{\eeqr}{\end{displaymath}}
\newcommand{\beqa}{\begin{eqnarray}}
\newcommand{\eeqa}{\end{eqnarray}}
\newcommand{\beqar}{\begin{eqnarray*}}
\newcommand{\eeqar}{\end{eqnarray*}}
\newcommand{\cN}{{\cal N}}

\newcommand{\cL}{{\cal L}}

\newcommand{\half}{\ensuremath{\frac{1}{2}}}

\newcommand{\N}[1]{\ensuremath{\cN=#1}}

\newcommand{\ps}{\ensuremath{\partial_\sigma}}

\newcommand{\cE}{{\cal E }}
\newcommand{\acosh}{\ensuremath{\mbox{arccosh}}}
\newcommand{\cotan}{\ensuremath{\mbox{cotan}}}

\title{\LARGE Single spike solutions for strings on $S^2$ and $S^3$}

\author{Riei Ishizeki, Martin Kruczenski \\
        Department of Physics, Purdue University, 525 Northwestern Avenue, \\
        W. Lafayette, IN 47907-2036. \\ 
E-mail: \email{rishizek@purdue.edu, markru@purdue.edu}}

\abstract{We study solutions for rigidly rotating strings on a two sphere. Among them we find two limiting cases that have a particular interest,
one is the already known giant magnon and the other we call the single spike solution. The limiting behavior of this last solution is a string 
infinitely wrapped around the equator. It differs from that solution by the existence of a single spike of height $\bar{\theta}$ that points 
toward the north pole.
 We study its properties and compute its energy $E$ and angular momentum $J$ as a function of $\bar{\theta}$. We further generalize the solution 
by adding one angular momentum to obtain a solution on $S^3$. We find a spin chain interpretation of these results in terms of free 
fermions and the Hubbard model but the exact relation with the same models derived from the field theory is not clear. 
}


\begin{document}

%%%% INTRODUCTION
\section{Introduction}
\label{intro}

 The idea of a large-N duality \cite{largeN} is a promising one for understanding the strong coupling limit of gauge theories.
The first example of a concrete duality in 3+1 dimensions was provided by the AdS/CFT correspondence \cite{malda} which conjectures
that \N{4} SYM in 3+1 dimensions is dual to IIB string theory on \adss{5}{5}. Strings states appear in the field theory \cite{bmn,GKP} 
as long gauge invariant operators. At this time, the string part of the correspondence is mostly understood at the classical level and
for that reason classical solutions play an important role in testing and unraveling the correspondence. Multi-spin solutions \cite{FT}
can be compared with particular field theory operators that can be represented as spin chains \cite{MZ}. The energy of the string agrees 
precisely with the conformal dimension of the corresponding operators \cite{BFST}. In fact one finds that spin chains are also connected to 
string theory \cite{kru,KRT,HL} by the fact that the classical action of a spin chain can be interpreted as the action of a string. In fact,
these spin chains are exactly the same as the ones arising from the field theory and can be used to derive the string sigma model, in the limit of 
a fast moving string, directly from the field theory \cite{kru}. 

 A different type of solution are those rotating in \ads{5} one of which is the spiky string \cite{spiky} which generalizes 
the rotating string of \cite{GKP} and describes higher twist operators from the field theory point of view. These spiky strings where 
generalized to the sphere in \cite{Ryang}.
 A particular limit of this solution, known as the giant magnon \cite{HM}, was identified with spin waves of short wavelength \cite{HM} opening
new possibilities and giving rise to various interesting results \cite{Dorey}--\cite{Bobev:2006fg}. 

 In a particular sector with $SU(2)$ symmetry the field theory description of the operators, at one loop in perturbation theory, 
is in terms of the ferromagnetic spin $\half$ Heisenberg model. In \cite{Rej:2005qt} it was conjectured that the Hubbard model was the
appropriate generalization at all couplings.  In this conjecture an important role is played by the antiferromagnetic state
\cite{Rej:2005qt,Zarembo:2005ur}. Later, in \cite{Roiban:2006jt} it was proposed that the antiferromagnetic state is described by a string 
wound around the equator a large number of times.
 One important point is that, when going from small to large coupling,  the Hubbard model interpolates between the Heisenberg model and free fermions.  
In this paper we find string solutions that look like an infinitely wound string with a spike pointing toward the north pole of the sphere. They 
should correspond to the free fermion states predicted by the Hubbard model which, at first sight, seems to be what we find. However, 
although the coefficients of the hopping term matches what the Hubbard model proposed from the field theory side, we require an extra term, 
proportional to the total fermion number, not present in the field theory side. Furthermore, analysis of the two angular momentum solution 
reveals a surprising dependence in the extra angular momentum which suggest an interpretation in terms of elementary excitations of fixed energy 
and not fixed angular momentum. In fact we can find a spin chain interpretation of the solutions in terms of the Hubbard model if we match the 
energy of the Hubbard model with the angular momentum of the string and the momentum in the Hubbard chain with the difference between energy and 
winding number in the string. Although these reproduces the string results, such identification is not the one resulting form the field theory 
calculation. The solutions therefore display the same rich behavior of the giant magnon but its direct relation to the field theory is unclear.    

 This paper is organized as follows. In the next section we describe briefly the spiky string and its T-dual in flat space. 
In section \ref{rigidstring} we propose and ansatz for a rigid string rotating on $S^2$ in static gauge and discuss the different possibilities according
to the values of the constants of motion. In section \ref{singlespike} we consider two limiting solutions, one is the giant magnon and the other
is the single spike. In section \ref{conformalgauge} we repeat the calculation in conformal gauge and discuss its relation to the sine-Gordon model.
 After that, in section \ref{J1J2}, we generalize the solution to a string rotating on $S^3$ with two angular momenta.  
In section \ref{spinchain} we discuss the interpretation of the results in terms of spin chains. We find a description in terms of free fermions
and in terms of the Hubbard model which, however, do not seem to be directly related to the field theory. Finally we give our conclusions in section 
\ref{conclusions}.

\section{Spiky strings in flat space and the T-dual solution}
\label{Tspiky}

 In \cite{spiky} the spiky string classical solution was introduced\footnote{In flat space similar solutions were known in the context of 
cosmic strings \cite{flatc}}. It is a rigidly rotating string with spikes as shown in fig.\ref{spikysol}.
It was generalized to the sphere in \cite{Ryang}. It can be described also in conformal gauge as shown in \cite{spikyS5} where a further generalization 
to $S^5$ was constructed in terms of solutions of the Neumann-Rosochatius system \cite{AFRT,ART,NR,Moser}. A limit of the spiky string, known as the 
giant magnon, is of special importance \cite{HM}. 

In this section we consider a T-dual solution to the flat space spiky string which, as we see, can be generalized
in precisely the same way. We start by the usual spiky string in flat space which, in conformal gauge is given by
\beqa
 x &=& A\, \cos\left((n-1)\ \sigma_+\right) + A\, (n-1)\, \cos\left(\sigma_-\right) \\
 y &=& A\, \sin\left((n-1)\ \sigma_+\right) + A\, (n-1)\, \sin\left(\sigma_-\right) \\
 t &=& 2\, A\, (n-1)\, \tau = A\, (n-1)\,(\sigma_++\sigma_-)
\label{flatsol}
\eeqa
 where $\sigma_+ = \tau+\sigma$, $\sigma_-=\tau-\sigma$, $n$ is an integer and $A$ is a constant which determines the size of the string. 
 It satisfies the equations of motion:
\beq
(\ps^2-\partial_\tau^2)X^i=\partial_{\sigma_+}\partial_{\sigma_-}X^i=0
\label{conf_eom}
\eeq
and the constraints
\beq
 \left(\partial_{\sigma_+} X\right)^2 = \left(\partial_{\sigma_-} X\right)^2 = 0 
\label{conf_const}
\eeq
 The cases of $n=3$ and $n=10$ are depicted in fig.\ref{spikysol}. 

\FIGURE{\epsfig{file=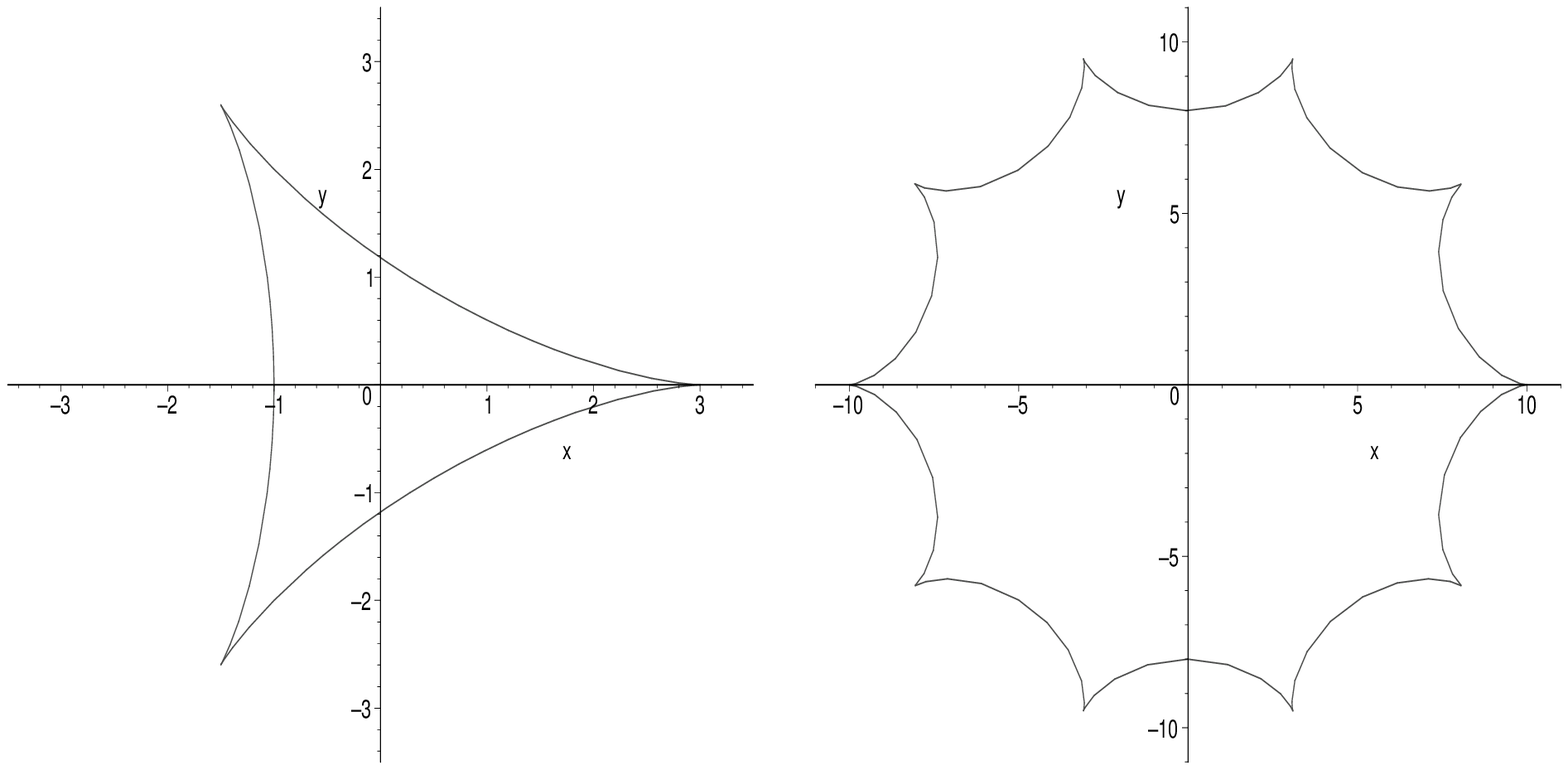, width=16cm}
\caption{Spiky strings in flat space. We show two examples with three and ten spikes respectively.}
\label{spikysol}
}

 Given a solution in flat space one can always construct a T-dual solution by changing the
sign of the left movers in one of the coordinates. Doing that we get a new solution:
\beqa
 x &=& A\, \cos\left((n-1)\ \sigma_+\right) - A\, (n-1)\, \cos\left(\sigma_-\right) \\
 y &=& A\, \sin\left((n-1)\ \sigma_+\right) + A\, (n-1)\, \sin\left(\sigma_-\right) \\
 t &=& 2\, A\, (n-1)\, \tau = A\, (n-1)\,(\sigma_++\sigma_-)
\label{Tflatsol}
\eeqa
that satisfies the same equations (\ref{conf_eom}) and constraints (\ref{conf_const}). The cases $n=3$ and $n=10$ are depicted in fig.\ref{Tspikysol}. 
 We intend to generalize these solutions to the sphere and consider a limit similar to the one that leads to the giant magnon.

\FIGURE{\epsfig{file=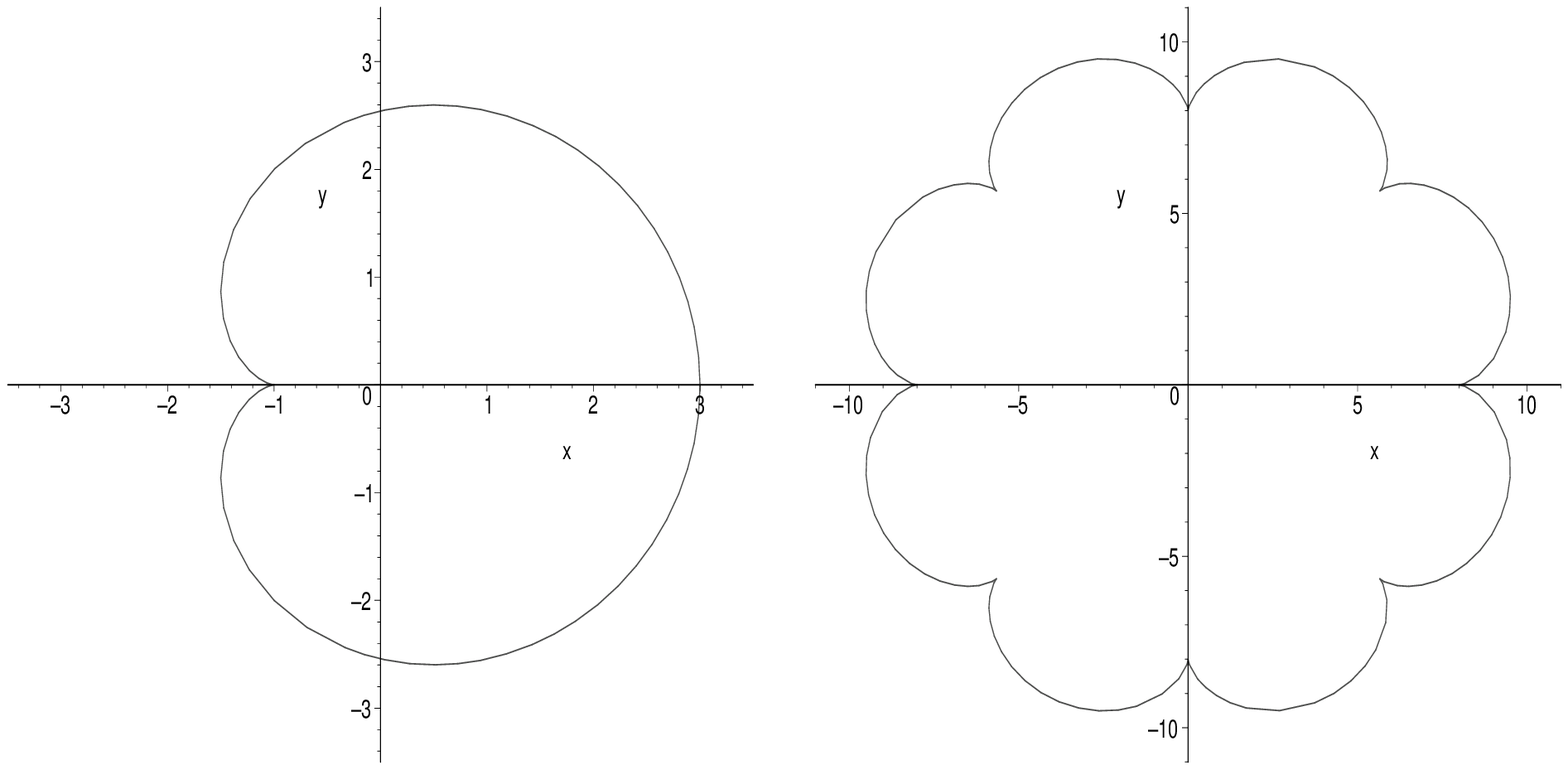, width=16cm}
\caption{T-duals of the solutions depicted in fig.\ref{spikysol}. The number of spikes changes by two when doing T-duality.}
\label{Tspikysol}
}

\section{Rigidly rotating strings on $S^2$}
\label{rigidstring}

The Nambu-Goto action is given by
\beq
 S = T \int d\sigma\, d\tau \cL 
   = T \int d\sigma\, d\tau \sqrt{(\partial_\sigma X.\partial_\tau X)^2 
                                -(\partial_\sigma X)^2 (\partial_\tau X)^2}
\label{SstNG}
\eeq
 where $T = \sqrt{\lambda}/2\pi$ is the string tension.
The space time metric is set as:
\beqa
 ds^2 &=& -dt^2 + d\theta^2+\sin^2\theta d\phi^2 \\
      &=&  G_{\mu\nu}(X) dX^\mu dX^\nu
\eeqa
We choose the parametrization 
\beqa
t&=&\kappa \tau \\
\phi&=& \omega\tau + \sigma \\
\theta&=&\theta(\sigma) 
\eeqa
Then, we can obtain the following equations of motion for (\ref{SstNG}),
\beqa
 \partial_\sigma \frac{\partial \cL}{\partial t'} + \partial_\tau \frac{\partial \cL}{\partial \dot{t}} 
 &=& \frac{\partial \cL}{\partial t} \label{eqt} \\
 \partial_\sigma \frac{\partial \cL}{\partial \phi'} + \partial_\tau \frac{\partial \cL}{\partial \dot{\phi}} 
 &=& \frac{\partial \cL}{\partial \phi} \label{eqphi} \\
 \partial_\sigma \frac{\partial \cL}{\partial \theta'} + \partial_\tau \frac{\partial \cL}{\partial \dot{\theta}} 
 &=& \frac{\partial \cL}{\partial \theta} \label{eqtheta} 
\eeqa
Solving (\ref{eqt}), we obtain,
\beq
  \theta' =  \frac{\kappa \sin \theta}{C} 
  \sqrt{\frac{\omega^2  \sin^2\theta - C^2}{\kappa^2 - \omega^2  \sin^2\theta}}
\label{theta1}
\eeq
where $C$ is a constant. We also obtain the same solution by solving 
(\ref{eqphi}), and, furthermore, the solution satisfies (\ref{eqtheta}).  Next, we 
compute the energy, angular momentum, and the difference in angle between 
two spikes.  

The energy is,
\beq
 E\ = 2 T \int_{\theta_0}^{\theta_1} \frac{d\theta}{\theta'} 
                                     \frac{\partial \cL}{\partial \dot{t}} 
    = 2 T \int_{\theta_0}^{\theta_1} d\theta \frac{\omega(C^2-\kappa^2)\sin\theta}
       {\kappa \sqrt{(\kappa^2-\omega^2\sin^2\theta)(\omega^2\sin^2\theta-C^2)}} 
\label{energy}
\eeq
the angular momentum is,
\beq
J\ = 2 T \int_{\theta_0}^{\theta_1} \frac{d\theta}{\theta'} \frac{\partial \cL}{\partial \dot{\phi}} 
    = 2 T \int_{\theta_0}^{\theta_1} d\theta \sin\theta \sqrt{\frac{\omega^2\sin^2\theta - C^2}{\kappa^2-\omega^2\sin^2\theta}} 
\label{ang_mom}
\eeq
and the difference in angle between two spikes is,
\beq 
 \Delta \phi = 2 \int \frac{d\theta}{\theta'} 
 = 2 \int_{\theta_0}^{\theta_1} d\theta \frac{C}{\kappa \sin \theta} \sqrt{\frac{\kappa^2-\omega^2\sin^2\theta}{\omega^2\sin^2\theta-C^2}}
 \label{ang_diff}
\eeq
 Requiring that the argument of the square root in (\ref{theta1}) be positive, we find that the 
range of $\theta$ can be, $C^2/\omega^2 < \sin^2\theta < \kappa^2/\omega^2$, 
or instead $\kappa^2/\omega^2 < \sin^2\theta < C^2/\omega^2$. Furthermore,
in the first case we can have (i) $\kappa^2/\omega^2<1$ or (ii) $\kappa^2/\omega^2>1$. In the second 
case we can have (iii) $C^2/\omega^2<1$ or (iv) $C^2/\omega^2>1$.  
 The different possibilities are summarized in table~\ref{table1}.

 In cases (i) and (ii) we can take the limit $|\omega|\rightarrow\kappa$ which gives rise to the giant magnon. In cases (iii) and (iv) 
we can take the limit $|\omega|\rightarrow|C|$ which gives rise to a solution we call the single spike and that we investigate in 
the rest of this paper.

\TABLE{
\begin{tabular}{|c|c|c|} \hline
	\ \ \ \  Case \ \ \ \ & \ \ \ \ \ Range \ \ \ \ \ & \ \ \ Conditions \ \ \ \\ \hline  \hline
	(i)  & \rule{0pt}{18pt} $\left(\frac{C}{\omega}\right)^2 < \sin^2\theta < 1 $ & $|C|<|\omega|<\kappa $\\ \hline
	(ii) & \rule{0pt}{18pt} $\left(\frac{C}{\omega}\right)^2 < \sin^2\theta < \left(\frac{\kappa}{\omega}\right)^2 $ 
                          & $|C|<\kappa<|\omega| $\\ \hline
       (iii) & \rule{0pt}{18pt} $\left(\frac{\kappa}{\omega}\right)^2 < \sin^2\theta < 1 $ & $\kappa<|\omega|<|C| $\\ \hline
        (iv) & \rule{0pt}{18pt} $\left(\frac{\kappa}{\omega}\right)^2  < \sin^2\theta < \left(\frac{C}{\omega}\right)^2 $ 
                                  & $\kappa<|C|<|\omega| $\\ \hline
\end{tabular}
\caption{Four cases for the motion of a rigid string on $S^2$ depending on the values of the constants. Notice that the cases 
$|\omega|<|C|<\kappa$ and $|\omega|<\kappa<|C|$ are forbidden. The limiting cases $|\omega|=|C|$ or $|\omega|=\kappa$ are studied separately.}
\label{table1}
}

\section{Limiting cases, giant magnon and single spike solution.}
\label{singlespike}

 In the two limiting cases, $|\omega|\rightarrow\kappa$ and $|\omega|\rightarrow|C|$ the equations simplify considerably and we can compute
the solution in terms of standard functions. We analyze both limits independently.

\subsection{First limiting case, giant magnon}

 Consider first case (ii) and define two angles:
\beq
  \theta_0 = \arcsin \frac{C}{\omega}, \ \ \ \ \ 
  \theta_1 = \arcsin \frac{\kappa}{\omega} ,
\label{t01a}
\eeq 
such that $\theta_0\le \theta\le \theta_1$. The limit $|\omega|\rightarrow\kappa$ corresponds to $\theta_1\rightarrow \frac{\pi}{2}$. In that
case we can integrate eq.(\ref{theta1}):
\beq
 \int \frac{\sin\theta_0\cos\theta\, d\theta}{\sin\theta\sqrt{\sin^2\theta-\sin^2\theta_0}} = \pm \, \sigma
\eeq
 and obtain:
\beq
 \pm \sigma = -\arcsin\left(\frac{\sin\theta_0}{\sin\theta}\right)
\eeq
or
\beq
 \sin\theta = \mp \frac{\sin\theta_0}{\sin\sigma}
\eeq
 Now, we have to evaluate equations (\ref{energy}), (\ref{ang_mom}), and (\ref{ang_diff})
in the limit, $\kappa \to \omega$.  From (\ref{ang_diff}), we obtain
\beq
 \Delta \phi = 2 \arccos \left(\frac{C}{\omega}\right) , \ \ \ \ \Rightarrow \ \ \ \ \frac{\Delta \phi}{2} = \frac{\pi}{2} - \theta_0
\eeq
 Equations, (\ref{energy}), and (\ref{ang_mom}), give a divergent value for the energy and angular momentum in this limit. 
The difference, however, is finite:
\beq
 E - J = 2 T\sin \frac{\Delta \phi}{2}
 = \frac{\sqrt{\lambda}}{\pi}\sin \frac{\Delta \phi}{2}
\eeq
 which are the known relations for the giant magnon. 

\subsection{Second limiting case, single spike}

 Consider now case (iv) in the limit $|\omega|\rightarrow C$. Again we define
\beq
  \theta_0 = \arcsin \frac{\kappa}{\omega}, \ \ \ \ \ 
  \theta_1 = \arcsin \frac{C}{\omega} ,
\eeq 
such that $\theta_0\le \theta\le \theta_1$. Notice that the definition is different from (\ref{t01a}) because the limits are interchanged.
The limit $|\omega|\rightarrow C$ now also corresponds to $\theta_1\rightarrow \frac{\pi}{2}$. Integrating eq.(\ref{theta1})
we get now:
\beq
 \frac{\cos^2\theta_0}{\sin\theta_0} \int \frac{\sin\theta\, d\theta}{\cos\theta\sqrt{\cos^2\theta_0-\cos^2\theta_0}} 
- \sin\theta_0 \int \frac{\cos\theta\, d\theta}{\sin\theta\sqrt{\sin^2\theta-\sin^2\theta_0}} = \pm\, \sigma
\eeq
which gives
\beq
\pm\, \sigma=
 \frac{\cos\theta_0}{\sin\theta_0} \acosh\left(\frac{\cos\theta_0}{\cos\theta}\right)-\arccos\left(\frac{\sin\theta_0}{\sin\theta}\right)
\eeq
In this case, we use (\ref{ang_mom}) first to obtain:
\beq
 J =  2T \frac{\kappa}{\omega} \sqrt{\left(\frac{\omega}{\kappa}\right)^2-1} = 2 T \cos\theta_0
\eeq
 On the other hand, (\ref{energy}) and (\ref{ang_diff}) diverge. 
By combining them, we obtain the finite result:
\beq
 E - T \Delta \phi = 2T\left(\frac{\pi}{2}-\theta_0\right)
\eeq
 The solution is plotted schematically in fig.\ref{sphere1}. We see that it is a string wrapped around the equator an infinite number of times
and with a single spike whose tip is at $\theta=\theta_0$. We define the height of the spike $\bar{\theta}$ as:
\beq
\bar{\theta} = \frac{\pi}{2} - \theta_0
\eeq
 In terms of the height we can rewrite:
\beqa
E - T \Delta \phi &=& \frac{\sqrt{\lambda}}{\pi}\, \bar{\theta} \\
J &=& \frac{\sqrt{\lambda}}{\pi}\sin \bar{\theta} 
\label{EJ1}
\eeqa
implying
\beq
\Delta = (E -T\Delta \phi) - J = \frac{\sqrt{\lambda}}{\pi} \left(\bar{\theta} - \sin \bar{\theta}\right)
\label{EJb}
\eeq
which is our final expression for the energy of the single spike solution on $S^2$. We define $\Delta$ by subtracting the ``free'' anomalous dimension
$J$ from the energy. We also need to subtract the winding number $T\Delta\phi$ to get a finite result. 

\FIGURE{\epsfig{file=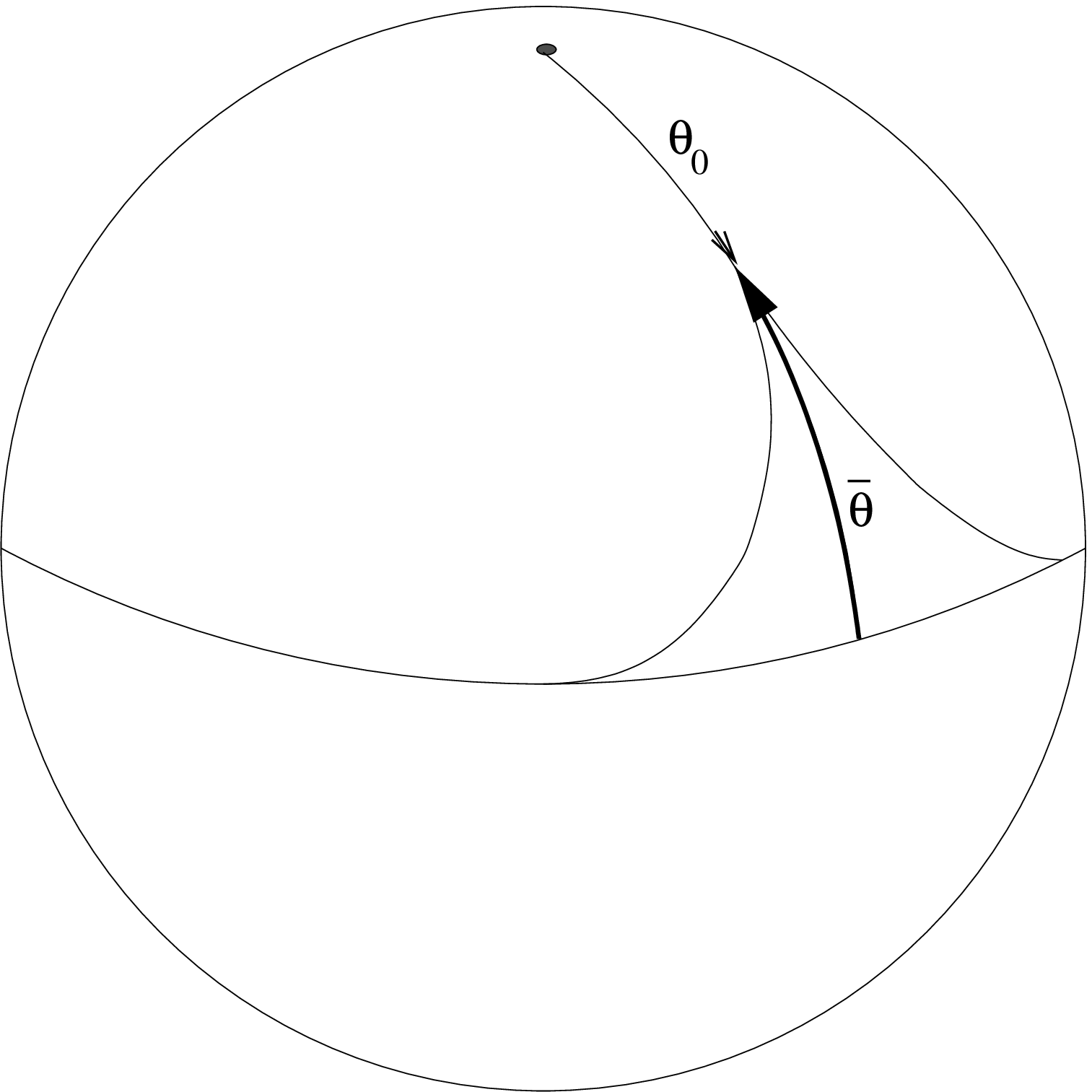, height=10cm}
\caption{Single spike solution.}
\label{sphere1}
}

\section{Conformal gauge and sine-Gordon model}
\label{conformalgauge}

 It is useful to obtain also the solution in conformal gauge. One simple way to do that is to use the results of \cite{spikyS5} where the 
spiky string was studied in conformal gauge. In fact, in that paper, after eq. (3.3) it is noted that two possibilities arise, which are described
as $\kappa=\omega_1$ and $\kappa=C_1$. In \cite{spikyS5} the first is shown to lead to the giant magnon. The second possibility actually leads to the 
single spike solution that we explore here. Let us summarize the points of \cite{spikyS5} that we need to use. A string moving on $S^5$ is described
in terms of three complex coordinates $X_{a=1\ldots 3}$ with metric:
\beq
 ds^2 = -dt^2 + \sum_a dX_a d\bar{X}_a ,\ \ \ \ \ \ \ \sum_a |X_a|^2 =1
\eeq
 Then the ansatz 
\beq
 X_a=x_a(\xi)e^{i\omega_a \tau}, \ \ \ \ \xi\equiv\alpha\sigma+\beta\tau,
\eeq
is proposed. The complex functions $x_a(\xi)$ are further parameterized as
\beq
 x_a(\xi)=r_a(\xi)e^{i\mu_a(\xi)},
\eeq
 where $r_a$, $\mu_a$ are real. The equations for the phases $\mu_a$ can be integrated giving
\beq
\mu'_a = \frac{1}{\alpha^2-\beta^2} \left[\frac{C_a}{r_a^2} + \beta\omega_a \right]
\eeq
 where $C_a$ are constants of motion. The equation of motion for the $r_a$ can then be derived from the Lagrangian 
\beq
\cL = \sum_a \left[(\alpha^2-\beta^2) r'_a{}^2-\frac{1}{\alpha^2-\beta^2}\frac{C_a^2}{r_a^2}-\frac{\alpha^2}{\alpha^2-\beta^2}\omega_a^2r_a^2 \right]
     +\Lambda\left(\sum_ar_a^2-a\right)
\eeq
where $\Lambda$ is a Lagrange multiplier. The Hamiltonian is 
\beq
H = \sum_a\left[(\alpha^2-\beta^2) r'_a{}^2+\frac{1}{\alpha^2-\beta^2}\frac{C_a^2}{r_a^2}+\frac{\alpha^2}{\alpha^2-\beta^2}\omega_a^2r_a^2\right]
\eeq
and the constraints are satisfied if
\beq
 \sum_a \omega_a C_a + \beta\kappa^2 =0 , \ \ \ \ \ H=\frac{\alpha^2+\beta^2}{\alpha^2-\beta^2}\kappa^2
\label{const1}
\eeq
 This is quite generic. We are interested in motion on an $S^2$ so we need to consider only the case $X_3=0$, $X_2$ real which implies
$\mu_2=0$ and $\omega_2=0$. Since we need $\mu'_2=0$, this also implies $C_2=0$. So the Hamiltonian reduces to
\beq
 H = (\alpha^2-\beta^2) \left(r'_1{}^2+r'_2{}^2\right)+ \frac{1}{\alpha^2-\beta^2}\frac{C_1^2}{r_1^2}+\frac{\alpha^2}{\alpha^2-\beta^2}\omega_1^2r_1^2
\eeq
 Since there is a constraint $r_1^2+r_2^2=1$ we can parameterize $r_{1,2}$ as 
\beq
r_1=\sin\theta, \ \ \ r_2=\cos\theta
\eeq
which gives
\beq
H = (\alpha^2-\beta^2) \theta'{}^2+\frac{C_1^2}{\alpha^2+\beta^2} \frac{1}{\sin^2\theta} + \frac{1}{\alpha^2-\beta^2}\omega_1^2 \sin^2\theta
\eeq
 Using conservation of energy and the constraints (\ref{const1}) we get
\beq
 \theta' = \pm\frac{\omega_1}{(\alpha^2-\beta^2)\sin\theta} \sqrt{(\sin^2\theta_0-\sin^2\theta)(\sin^2\theta-\sin^2\theta_1)}
\eeq
where
\beq
\sin^2\theta_0 = \frac{\beta^2\kappa^2}{\omega^2}, \ \ \ \ \ \ \sin^2\theta_1 = \frac{\kappa^2}{\omega^2}
\eeq
We see that two possibilities arise: $\theta_0<\theta<\theta_1$ or $\theta_1<\theta<\theta_0$. In the limit $\omega\to \kappa $ we have 
$\theta_1=\frac{\pi}{2}$, and in the limit $|\omega|\to |C_1|$ we have $\theta_0=\frac{\pi}{2}$. Both cases can be integrated as before.
We get for the giant magnon
\beqa
 \cos\theta &=& \frac{\cos\theta_0}{\cosh\xi} \\
 \mu_1 &=& - \frac{\cos\theta_0}{\sin\theta_0}\arctan\left(\frac{\cos\theta_0}{\sin\theta_0}\tanh\xi\right) \\
 \phi &=& \mu_1 + \omega_1\tau \\
 \xi &=& \sigma+ \sin\theta_0 \tau \ \ \ \ \ \omega=\kappa=\cos\theta_0
\eeqa
 where we chose $\alpha$ and $\beta$ in an appropriate way to simplify the solution and $\theta,\phi$ are the angles parameterizing the $S^2$ . 
For the single spike solution we have
\beqa
\cos\theta &=& \frac{\cos\theta_1}{\cosh\xi} \\
 \mu_1 &=& \frac{\cos\theta_1}{\sin\theta_1} \xi - \arctan\left(\frac{\cos\theta_1}{\sin\theta_1}\tanh\xi\right) \\
 \phi &=& \mu_1 + \omega_1\tau  \\
 \xi &=& \sigma\sin\theta_1 + \tau, \ \ \  \omega=-\cotan\theta_1,  \ \ \ \kappa = \cos\theta_1
\eeqa
 where we now chose $\beta=1$. Notice that for the giant magnon $\beta<\alpha$ whereas for the single spike $\alpha<\beta$. 
If we compute the energy, angular momentum and $\Delta\phi$ we get the same result as before.

 As in \cite{HM} we can see a relation to the sine-Gordon model. In conformal gauge it is interesting to compute the determinant 
of the world-sheet metric given by
\beqa
 h &=& \left(-\kappa^2+\dot{\theta}^2 + \sin^2\theta \dot{\phi}^2\right)\left(\theta'{}^2 + \sin^2\theta \phi'{}^2\right)
     -\left(\dot{\theta}\theta'+\sin^2\theta\dot{\phi}\phi'\right)^2 \\
   &=& - \left(\theta'{}^2 + \sin^2\theta \phi'{}^2\right)^2
\eeqa
 where we used the conformal constraints. Replacing the solutions we obtain:
\beqa
 \frac{1}{\kappa^2} \sqrt{-h} &=& \frac{1}{\cosh^2\xi}, \ \ \ \ \mbox{giant magnon} \\ \\
 \frac{1}{\kappa^2} \sqrt{-h} &=& 1-\frac{1}{\cosh^2\xi}, \ \ \ \ \mbox{single spike} 
\eeqa 
 If we now define an angle through:
\beq
 \sin^2 \Phi =  \frac{1}{\kappa^2} \sqrt{-h} 
\eeq
we get
\beqa
 \Phi &=&\arcsin \left(\frac{1}{\cosh\xi}\right), \ \ \ \ \mbox{giant magnon}  \\ \\
 \Phi &=&\arcsin  \tanh \xi , \ \ \ \ \ \ \ \ \ \mbox{single spike} 
\eeqa
 It is now easy to check that both angles satisfy the sine-Gordon equation:
\beq
 \left(\partial_\tau^2-\partial_\sigma^2\right) \Phi + \half\kappa^2\sin(2\Phi)=0
\label{sG}
\eeq
 Note however that they correspond to different potentials, since $\Phi(\pm\infty)=k\pi$ for the giant magnon and $\Phi(\pm\infty)=\frac{\pi}{2}+k\phi$
for the single spike. Other way of saying it, is that $\sigma$ and $\tau$ are interchanged when going from one to the other solution (because 
interchanging $\sigma$ and $\tau$ is equivalent to changing the sign of the potential in (\ref{sG})). In any case, the
relation to sine-Gordon should be useful to compute scattering of single spikes as in \cite{HM} although we do not pursue this here. Similarly,
scattering of spikes in the spiky string \cite{spiky} is determined by the sinh-Gordon model.

\section{Strings on $S^3$, two angular momenta}
\label{J1J2}

In a previous section, we analyzed the motion of the string on a two sphere. 
In this section  we add one more 
dimension to the sphere, and investigate the motion with an extra angular 
momentum.

We can use the same action (\ref{SstNG}) as in the previous section. 
The space time metric is now:
\beqa
 ds^2 &=& -dt^2 +  d\theta^2+\sin^2\theta d\phi_1^2+\cos^2\theta d\phi_2^2 \\
      &=&  G_{\mu\nu}(X) dX^\mu dX^\nu
\eeqa
This time, we choose the parametrization as 
\beqa
t&=&\kappa \tau \\
\theta&=&\theta(\sigma) \\ 
\phi_1&=&\omega_1\tau + \sigma \\
\phi_2&=&\phi_2(\sigma) + \omega_2\tau 
\eeqa
Then, we can obtain the following equations of motion for (\ref{SstNG}),
\beqa
 \partial_\sigma \frac{\partial \cL}{\partial t'} + \partial_\tau \frac{\partial \cL}{\partial \dot{t}} &=& \frac{\partial \cL}{\partial t} \label{eqt2} \\
 \partial_\sigma \frac{\partial \cL}{\partial \theta'} + \partial_\tau \frac{\partial \cL}{\partial \dot{\theta}} &=& \frac{\partial \cL}{\partial \theta} \label{eqtheta2} \\ 
 \partial_\sigma \frac{\partial \cL}{\partial \phi_1'} + \partial_\tau \frac{\partial \cL}{\partial \dot{\phi_1}} &=& \frac{\partial \cL}{\partial \phi_1} \label{eqphi12} \\
 \partial_\sigma \frac{\partial \cL}{\partial \phi_2'} + \partial_\tau \frac{\partial \cL}{\partial \dot{\phi_2}} &=& \frac{\partial \cL}{\partial \phi_2} \label{eqphi22} 
\eeqa
Solving (\ref{eqt2}) and (\ref{eqphi12}) for $\phi_2'$, we obtain
\beq
\phi_2' = \frac{\sin^2\theta \left\{\kappa(\kappa C_1 -\omega_1 C_2)-\omega_2^2 C_1\cos^2 \theta \right\}}{\omega_2 \cos^2 \theta (\kappa C_2 -\omega_1 C_1 \sin^2 \theta)}
\eeq
where $C_1$ and $C_2$ are constants. The equation for $\theta'$ is 
rather complicated so, before considering it we choose the constants of motion appropriately such that 
\beq
\theta' \to 0 \quad \textrm{as} \quad \theta \to \frac{\pi}{2}
\eeq
 We obtain the following values of the constants:
\beq
C_1 = \omega_1 \quad \textrm{and} \quad C_2 = \kappa
\eeq
Now, with these values we obtain simpler equations:
\beq
\phi_2' = \frac{\omega_1\omega_2\sin^2\theta}{\omega_1^2\sin^2\theta-\kappa^2} 
\label{phi2} 
\eeq
\beq
\theta' = \frac{\kappa \sin \theta \cos\theta}{\omega_1^2\sin^2\theta-\kappa^2}
          \sqrt{(\omega_1^2-\omega_2^2)\sin^2\theta - \kappa^2}
\label{theta2} 
\eeq
As in the previous section, we compute the energy, two angular
momenta, and the difference in angle between the two endpoints of the string 
using eqs. (\ref{phi2}) and (\ref{theta2}).

The energy is,
\beq
 E\ = 2T \int_{\theta_0}^{\theta_1} \frac{d\theta}{\theta'} \frac{\partial \cL}{\partial \dot{t}} 
 = 2 T \int_{\theta_0}^{\theta_1} d\theta \frac{(\omega_1^2-\kappa^2)\sin\theta}{\kappa \cos\theta \sqrt{(\omega_1^2-\omega_2^2)\sin^2\theta - \kappa^2}} 
\label{energy2}
\eeq
the first angular momentum is,
\beq
 J_1\ = 2T \int_{\theta_0}^{\theta_1} \frac{d\theta}{\theta'} \frac{\partial \cL}{\partial \dot{\phi_1}}
 = 2 T \int_{\theta_0}^{\theta_1} d\theta \frac{\omega_1\sin\theta\cos\theta}
 {\sqrt{(\omega_1^2-\omega_2^2)\sin^2\theta - \kappa^2}} 
\label{ang_mom21}
\eeq
the second angular momentum is,
\beq
 J_2\ = 2T \int_{\theta_0}^{\theta_1} \frac{d\theta}{\theta'} \frac{\partial \cL}{\partial \dot{\phi_2}}
 = 2 T \int_{\theta_0}^{\theta_1} d\theta \frac{\omega_2\sin\theta\cos\theta}
 {\sqrt{(\omega_1^2-\omega_2^2)\sin^2\theta - \kappa^2}} 
\label{ang_mom22}
\eeq
and the difference in angle between the two endpoints of the strings is,
\beq 
 \Delta \phi = 2 \int_{\theta_0}^{\theta_1} \frac{d\theta}{\theta'} = 2 \int_{\theta_0}^{\theta_1} d\theta \frac{\omega_1^2\sin^2\theta-\kappa^2}
 {\kappa\sin\theta\cos\theta \sqrt{(\omega_1^2-\omega_2^2)\sin^2\theta - \kappa^2}} 
 \label{ang_diff2}
\eeq
where $\theta_0=\arcsin\left(\kappa/\sqrt{\omega_1^2-\omega_2^2}\right)$ with 
$\omega_1^2>\omega_2^2$, and $\theta_1=\pi/2$. Here, $\theta_0$ is chosen such that 
the inside square root of (\ref{theta2}) is positive.  Since 
$\arcsin(\kappa/\omega_1) < \arcsin\left(\kappa/\sqrt{\omega_1^2-\omega_2^2}\right) < \pi/2$,
$\theta$ can never reach a value such that $\sin\theta=\kappa/\omega_1$.  Thus, in
this case, $\theta'$ does not go to infinity at any point. Also, we remind the reader that the tension $T$ is
$T=\frac{\sqrt{\lambda}}{2\pi}$. 

 Doing the integrals we find the following results:
\beqa
E-T\Delta \phi &=& 2T \bar{\theta} \\ 
J_1 &=& 2T \frac{1}{\cos\gamma} \sin\bar{\theta} \\
J_2 &=& 2T \frac{\sin\gamma}{\cos\gamma} \sin\bar{\theta}
\eeqa
where we defined
\beq
 \bar{\theta}=\frac{\pi}{2}-\theta_0, \ \ \ \  
 \sin\theta_0 = \frac{\kappa}{\sqrt{\omega_1^2-\omega_2^2}} , \ \ \ \ \ \sin\gamma = \frac{\omega_1}{\omega_2}
\eeq
 The result can also be written as:
\beqa
 E-T\Delta \phi &=& \frac{\sqrt{\lambda}}{\pi}\, \bar{\theta} \\ 
 J_1 &=& \sqrt{J_2^2+\frac{\lambda}{\pi^2}\sin^2\bar{\theta}} 
\label{EJ2}
\eeqa
 In the case of $J_2=0$ we recover the expressions (\ref{EJ1}) for the energy and angular momentum of the single spike on $S^2$. 

\section{Spin chain interpretation}
\label{spinchain}

 It has become standard that, from the string solutions, only the parts of the string that move almost at the speed of light have
a simple interpretation in the field theory. This was valid in the BMN case \cite{bmn} and also when mapping the string
and spin chain actions \cite{kru} or in the spiky strings \cite{spiky}. The case of the giant magnon \cite{HM} might seem different but in fact,
in such solution, the sigma coordinate spans an infinite range, most of which corresponds to the string close to the equator moving
at the speed of light. In our case, the only part of the string moving at the speed of light is the spike which one might think could
be interpreted as in the case of the spiky string. In principle, one might hope for more because the infinitely wound string was associated with 
the anti-ferromagnetic state in \cite{Roiban:2006jt} and our solution is a perturbation of that. However it is not completely clear to us how the
relation to the anti-ferromagnetic state works so, in this paper, we do not pursue this avenue any further. 

 Summarizing, given the fact that the string does not move at the speed of light, we anticipate a difficulty in mapping the solutions to the field theory.
On the other hand, it turns out to be rather simple to find a spin chain interpretation of the results, if, at this stage, we do not require
a field theory derivation of such spin chain. In fact, if we look at the string with two angular momenta whose energy and angular momentum 
are given by (\ref{EJ2}):
\beqa
 E-T\Delta \phi &=& \frac{\sqrt{\lambda}}{\pi}\, \bar{\theta} \\ 
 J_1 &=& \sqrt{J_2^2+\frac{\lambda}{\pi^2}\sin^2\bar{\theta}} 
\eeqa
 we find that it is very similar to the giant magnon result \cite{HM,Dorey,Dorey2} if we interpret $\bar{\theta}$ as half the momentum of the magnon 
$\bar{\theta}=\half p$.
The main difference is that we should interpret the energy of the magnon as $J_1$ and not $E$ which is rather surprising. Also we have an extra quantity,
namely $E-T\Delta\phi$ that we computed and should be interpreted as the momentum of the magnon. Therefore, a spin chain interpretation of the result
is to take a Hubbard chain and identify the Hamiltonian of the chain with the angular momentum of the string and the momentum in the chain with the 
difference $E-T\Delta\phi$ in the string. This can be loosely understood as a $\sigma$ , $\tau$ interchange which is what we also saw in the 
relation to the sine-Gordon model.
 Although this allows us a spin chain interpretation of the solutions and shows that their dynamics is as rich as the one of the rotating 
strings, the problem is that we did not derive the spin chain from the field theory. We leave the important problem of mapping these solutions to 
operators in \N{4} SYM theory for the future. An important clue might be the ideas of \cite{Roiban:2006jt}.

\subsection{Fermi sea}

 In this subsection we point a curious fact about the solution on $S^2$ that suggests a different but related spin chain interpretation. Since it does not
generalize to the $S^3$ case we believe to be an interesting but very particular interpretation. 

 In \cite{Rej:2005qt} the Hubbard model was proposed as a way to determine the conformal dimension of various operators in the field theory. 
The Hamiltonian is given by
\beq
 H = -t\sum_{i=-\infty\ldots\infty,\alpha=\uparrow,\downarrow}\left( c_{i,\alpha}^\dagger c_{i+1,\alpha}+c^\dagger_{i+1,\alpha}c_{i,\alpha}\right) 
     + U\sum_{i=-\infty\ldots\infty} c^{\dagger}_{i\uparrow}c_{i\uparrow} c^\dagger_{i\downarrow} c_{i\downarrow}
\eeq
 where
\beq
 U =-\half, \ \ \ t=-\frac{\sqrt{\lambda}}{8\pi}
\eeq
 The Hamiltonian corresponds to a system of electrons living in a one dimensional lattice whose sites are labeled by the index $i$. There is a
hopping term proportional to $t$ and an on-site repulsion modeling the Coulomb repulsion. The electron has two states, spin up and spin down 
labeled by the index $\alpha=\uparrow,\downarrow$. 
In the limit of small coupling $\lambda\ll1$ the second term can be ignored and we get a system of free fermions. 
The energy of a single fermion is given in terms of its momentum by
\beq
 \epsilon(k) = -2t\cos k 
\eeq
 We see that there are fermionic states with negative energy so the ground state is half-filled. It turns out that for our purpose it is necessary
to add a chemical potential $\mu$ such that the ground state is empty. We take then the Hamiltonian to be
\beq
 \tilde{H} = -t\sum_{i=-\infty\ldots\infty,\alpha=\uparrow,\downarrow}\left( c_{i,\alpha}^\dagger c_{i+1,\alpha}+c^\dagger_{i+1,\alpha}c_{i,\alpha}
          + 2 c_{i,\alpha}^\dagger c_{i,\alpha}  \right) 
     + U\sum_{i=-\infty\ldots\infty} c^{\dagger}_{i\uparrow}c_{i\uparrow} c^\dagger_{i\downarrow} c_{i\downarrow}
\eeq
namely, $\mu=2t$ if we take $\tilde{H} =H-\mu N$ with $N$ the number of fermions. The new single particle energy is now
\beq
 \epsilon(k) = -2t (1+\cos k) 
\eeq
 which is always positive (since $t<0$). It is also convenient to define $\kappa=\pi-k$ since the lowest energy state has $k=\pi$. We get
\beq
 \epsilon(\kappa) = -2t (1-\cos \kappa) 
\eeq
 If we now fill the states up to some Fermi momentum $\kappa_F$ the total energy is given by
\beq
 \cE = -8t \int_0^{\kappa_F}\!\! (1-\cos \kappa)\,d\kappa = - 8 t (\kappa_F -\sin \kappa_F) = \frac{\sqrt{\lambda}}{\pi} (\kappa_F -\sin \kappa_F)  
\eeq
where we introduced a degeneracy factor of four, two because the states with momentum $\kappa$ and $-\kappa$ have the same energy and another two because
of the spin degeneracy. If we identify $\bar{\theta}=\kappa_F$, and $\Delta=\cE$ then we get precisely eq.(\ref{EJb}). Tantalizingly it does so up to the 
coefficient.  Notice that, in the previous subsection, we identified $\theta=\half p$ where $p$ is the momentum of the magnon.
However, in this model a magnon is a bound state of two fermions, each with momentum $\half p$. It seems then that, in both cases, $\bar{\theta}$
is related to some underlying fermionic momentum. Filling the Fermi sea corresponds to wrapping the string once more around the equator since the
energy changes by $2\pi T$ when $\kappa_F=\pi$. Unfortunately, as mentioned before the interpretation in terms of a Fermi sea does not appear to 
generalize to the solution with two angular momenta and therefore we regard this interpretation as a curiosity. We mentioned it here because it 
might be useful for other purposes and also was the first interpretation we found. 
 
\section{Conclusions} 
\label{conclusions}

 In this paper we find and study new solutions for rigid strings moving on a sphere. They asymptote to a solution infinitely wrapped around the equator 
and at rest. Therefore are fundamentally different from the rotating strings where the string moves close to the speed of light. They 
belong to the category of slowly moving strings described in \cite{Roiban:2006jt}. Nevertheless we find that they have an interesting and rich 
dynamics that seems to be described by the same spin chains as their rotating strings counterparts. For example, the Hubbard model proposed 
in \cite{Rej:2005qt} that describes the two angular momentum giant magnon also appears useful to describe these solutions. The main difference 
is that we should map the angular momentum 
of the string to the energy of the spin chain and the difference between energy and wrapping number to the momentum of the spin chain. This makes
difficult the interpretation of the spin chain in the field theory. We leave this last part for future and presumably difficult work. However other 
type of work seems more directly accessible, for example it would be interesting to study the scattering of single spikes and compare them 
with the scattering of magnons in the spin chain. This would strengthen the map we propose between these solutions and the spin chains. We hope
to report on this in the near future. Also, related ideas can be found in the more recent work \cite{recent}.

\section{Acknowledgments}
We are grateful to T. ter Veldhuis and Chi Xiong for comments and discussions. The work of R.I was supported in part by 
the Purdue Research Foundation (PRF) and the one of M.K. was supported in part by NSF under grant PHY-0653357.

\section*{Note added}
 While this work was being completed we learned about related work on the T-dual solutions of the spiky strings by 
A. Mosaffa and B. Safarzadeh \cite{Mosaffa:2007ty}.


\begin{thebibliography}{99}        

\bibitem{largeN}
G.~'t Hooft,
   ``A Two-Dimensional Model For Mesons,''
  Nucl.\ Phys.\ B {\bf 75}, 461 (1974).
  %%CITATION = NUPHA,B75,461;%%

\bibitem{malda}
J.~Maldacena,
``The large $N$ limit of superconformal field theories and supergravity,''
Adv.\ Theor.\ Math.\ Phys.\  {\bf 2}, 231 (1998)
[Int.\ J.\ Theor.\ Phys.\  {\bf 38}, 1113 (1998)],
{\tt hep-th/9711200}, \\
%%CITATION = HEP-TH 9711200;%%
%\cite{Gubser:1998bc}
%\bibitem{Gubser:1998bc}
S.~S.~Gubser, I.~R.~Klebanov and A.~M.~Polyakov,
``Gauge theory correlators from non-critical string theory,''
Phys.\ Lett.\ B {\bf 428}, 105 (1998)
[arXiv:hep-th/9802109], \\
%%CITATION = HEP-TH 9802109;%%
%\cite{Witten:1998qj}
%\bibitem{Witten:1998qj}
E.~Witten,
``Anti-de Sitter space and holography,''
Adv.\ Theor.\ Math.\ Phys.\  {\bf 2}, 253 (1998)
[arXiv:hep-th/9802150], \\
%%CITATION = HEP-TH 9802150;%%

  
%\cite{Berenstein:2002jq}
\bibitem{bmn}
D.~Berenstein, J.~M.~Maldacena and H.~Nastase,
``Strings in flat space and pp waves from N = 4 super Yang Mills,''
JHEP {\bf 0204}, 013 (2002)
[arXiv:hep-th/0202021].
%%CITATION = HEP-TH 0202021;%%
%\cite{Gubser:2002tv}

\bibitem{GKP}
S.~S.~Gubser, I.~R.~Klebanov and A.~M.~Polyakov,
``A semi-classical limit of the gauge/string correspondence,''
Nucl.\ Phys.\ B {\bf 636}, 99 (2002)
[arXiv:hep-th/0204051].
%%CITATION = HEP-TH 0204051;%%

%\cite{Frolov:2003xy}
\bibitem{FT}
S.~Frolov and A.~A.~Tseytlin,
  ``Multi-spin string solutions in AdS(5) x S**5,''
  Nucl.\ Phys.\ B {\bf 668}, 77 (2003)
  [arXiv:hep-th/0304255].
  %%CITATION = HEP-TH 0304255;%%
``Rotating string solutions: AdS/CFT duality in non-supersymmetric  sectors,''
Phys.\ Lett.\ B {\bf 570}, 96 (2003)
[arXiv:hep-th/0306143].
%%CITATION = HEP-TH 0306143;%%

%\cite{Minahan:2002ve}
\bibitem{MZ}
J.~A.~Minahan and K.~Zarembo,
``The Bethe-ansatz for N = 4 super Yang-Mills,''
JHEP {\bf 0303} (2003) 013
[arXiv:hep-th/0212208].
%%CITATION = HEP-TH 0212208;%%

%\cite{Beisert:2003ea}
\bibitem{BFST}
N.~Beisert, J.~A.~Minahan, M.~Staudacher and K.~Zarembo,
``Stringing spins and spinning strings,''
JHEP {\bf 0309}, 010 (2003)
[arXiv:hep-th/0306139], \\
%%CITATION = HEP-TH 0306139;%%
N.~Beisert, S.~Frolov, M.~Staudacher and A.~A.~Tseytlin,
``Precision spectroscopy of AdS/CFT,''
JHEP {\bf 0310}, 037 (2003)
[arXiv:hep-th/0308117].
%%CITATION = HEP-TH 0308117;%%


%\cite{Kruczenski:2003gt}
\bibitem{kru}
M.~Kruczenski,
``Spin chains and string theory,'', 
Phy. Rev. Lett {\bf 93}, 161602 (2004),
[arXiv:hep-th/0311203].
%%CITATION = HEP-TH 0311203;%%

%\cite{Kruczenski:2004kw}
\bibitem{KRT}
M.~Kruczenski, A.~V.~Ryzhov and A.~A.~Tseytlin,
``Large spin limit of AdS(5) x S**5 string theory and low energy expansion of
ferromagnetic spin chains,''
Nucl.\ Phys.\ B {\bf 692}, 3 (2004)
[arXiv:hep-th/0403120].
%%CITATION = HEP-TH 0403120;%%

%\cite{Hernandez:2004uw}
\bibitem{HL}
R.~Hernandez and E.~Lopez,
``The SU(3) spin chain sigma model and string theory,''
JHEP {\bf 0404}, 052 (2004)
[arXiv:hep-th/0403139].
%%CITATION = HEP-TH 0403139;%%

%\cite{Kruczenski:2004wg}
\bibitem{spiky}
  M.~Kruczenski,
   ``Spiky strings and single trace operators in gauge theories,''
  JHEP {\bf 0508}, 014 (2005)
  [arXiv:hep-th/0410226].
  %%CITATION = HEP-TH 0410226;%%

%\cite{Ryang:2005yd}
\bibitem{Ryang}
  S.~Ryang,
   ``Wound and rotating strings in AdS(5) x S**5,''
  JHEP {\bf 0508}, 047 (2005)
  [arXiv:hep-th/0503239].
  %%CITATION = HEP-TH 0503239;%%

%\cite{Hofman:2006xt}
\bibitem{HM}
  D.~M.~Hofman and J.~M.~Maldacena,
   ``Giant magnons,''
  arXiv:hep-th/0604135.
  %%CITATION = HEP-TH 0604135;%%

%%% giant magnon


%\cite{Dorey:2006dq}
\bibitem{Dorey}
  N.~Dorey,
  ``Magnon bound states and the AdS/CFT correspondence,''
  arXiv:hep-th/0604175.
  %%CITATION = HEP-TH 0604175;%%
  
  \bibitem{Dorey2}
  H.~Y.~Chen, N.~Dorey and K.~Okamura,
  ``Dyonic giant magnons,''
  arXiv:hep-th/0605155.
  %%CITATION = HEP-TH 0605155;%%

%\cite{Astolfi:2007uz}
\bibitem{Astolfi:2007uz}
  D.~Astolfi, V.~Forini, G.~Grignani and G.~W.~Semenoff,
  ``Gauge invariant finite size spectrum of the giant magnon,''
  arXiv:hep-th/0702043.
  %%CITATION = HEP-TH/0702043;%%

%\cite{Minahan:2007gf}
\bibitem{Minahan:2007gf}
  J.~A.~Minahan,
  ``Zero modes for the giant magnon,''
  JHEP {\bf 0702}, 048 (2007)
  [arXiv:hep-th/0701005].
  %%CITATION = JHEPA,0702,048;%%

%\cite{Hatsuda:2006ty}
\bibitem{Hatsuda:2006ty}
  Y.~Hatsuda and K.~Okamura,
  ``Emergent classical strings from matrix model,''
  arXiv:hep-th/0612269.
  %%CITATION = HEP-TH/0612269;%%

%\cite{Bozhilov:2006gh}
\bibitem{Bozhilov:2006gh}
  P.~Bozhilov,
  ``A note on two-spin magnon-like energy-charge relations from M-theory
  %viewpoint,''
  arXiv:hep-th/0612175,\\
  P.~Bozhilov,
  ``Neumann and Neumann-Rosochatius integrable systems from membranes on
  $AdS_4xS^7$,''
  arXiv:0704.3082 [hep-th].
  %%CITATION = ARXIV:0704.3082;%%

%\cite{Arutyunov:2006gs}
\bibitem{AFZ}
  G.~Arutyunov, S.~Frolov and M.~Zamaklar,
   ``Finite-size effects from giant magnons,''
arXiv:hep-th/0606126.
 %%CITATION = HEP-TH/0606126;%%

%\cite{Maldacena:2006rv}
\bibitem{Maldacena:2006rv}
  J.~Maldacena and I.~Swanson,
  ``Connecting giant magnons to the pp-wave: An interpolating limit of AdS(5) x
  S**5,''
  arXiv:hep-th/0612079.
  %%CITATION = HEP-TH/0612079;%%

%\cite{Kalousios:2006xy}
\bibitem{Kalousios:2006xy}
  C.~Kalousios, M.~Spradlin and A.~Volovich,
  ``Dressing the giant magnon. II,''
  arXiv:hep-th/0611033.
  %%CITATION = HEP-TH/0611033;%%

%\cite{Itoyama:2006cg}
\bibitem{Itoyama:2006cg}
  H.~Itoyama and T.~Oota,
  ``The AdS(5) x S**5 superstrings in the generalized light-cone gauge,''
  arXiv:hep-th/0610325.
  %%CITATION = HEP-TH/0610325;%%

%\cite{Ryang:2006yq}
\bibitem{Ryang:2006yq}
  S.~Ryang,
  ``Three-spin giant magnons in AdS(5) x S**5,''
  JHEP {\bf 0612}, 043 (2006)
  [arXiv:hep-th/0610037].
  %%CITATION = JHEPA,0612,043;%%

%\cite{Hirano:2006ti}
\bibitem{Hirano:2006ti}
  S.~Hirano,
  ``Fat magnon,''
  arXiv:hep-th/0610027.
  %%CITATION = HEP-TH/0610027;%%

%\cite{Okamura:2006zv}
\bibitem{Okamura:2006zv}
  K.~Okamura and R.~Suzuki,
  ``A perspective on classical strings from complex sine-Gordon solitons,''
  Phys.\ Rev.\  D {\bf 75}, 046001 (2007)
  [arXiv:hep-th/0609026].
  %%CITATION = PHRVA,D75,046001;%%

%\cite{Roiban:2006gs}
\bibitem{Roiban:2006gs}
  R.~Roiban,
  ``Magnon bound-state scattering in gauge and string theory,''
  arXiv:hep-th/0608049.
  %%CITATION = HEP-TH/0608049;%%

%\cite{Chen:2006gq}
\bibitem{Chen:2006gq}
  H.~Y.~Chen, N.~Dorey and K.~Okamura,
  ``On the scattering of magnon boundstates,''
  JHEP {\bf 0611}, 035 (2006)
  [arXiv:hep-th/0608047].
  %%CITATION = JHEPA,0611,035;%%

%\cite{Gomez:2006va}
\bibitem{Gomez:2006va}
  C.~Gomez and R.~Hernandez,
  ``The magnon kinematics of the AdS/CFT correspondence,''
  JHEP {\bf 0611}, 021 (2006)
  [arXiv:hep-th/0608029].
  %%CITATION = JHEPA,0611,021;%%

\bibitem{spradlin} 
 M.~Spradlin and A.~Volovich,
   ``Dressing the Giant Magnon,''
  arXiv:hep-th/0607009.
  %%CITATION = HEP-TH 0607009;%%

%\cite{Beccaria:2006td}
\bibitem{Beccaria:2006td}
  M.~Beccaria and L.~Del Debbio,
  ``Bethe Ansatz solutions for highest states in N = 4 SYM and AdS/CFT
  duality,''
  JHEP {\bf 0609}, 025 (2006)
  [arXiv:hep-th/0607236].
  %%CITATION = JHEPA,0609,025;%%

%\cite{Vazquez:2006hd}
\bibitem{Vazquez:2006hd}
  S.~E.~Vazquez,
 ``BPS condensates, matrix models and emergent string theory,''
  JHEP {\bf 0701}, 101 (2007)
  [arXiv:hep-th/0607204].
  %%CITATION = JHEPA,0701,101;%%

%\cite{Huang:2006vz}
\bibitem{Huang:2006vz}
  W.~H.~Huang,
 ``Giant magnons under NS-NS and Melvin fields,''
  JHEP {\bf 0612}, 040 (2006)
  [arXiv:hep-th/0607161].
  %%CITATION = JHEPA,0612,040;%%

%\cite{Bozhilov:2006bi}
\bibitem{Bozhilov:2006bi}
  P.~Bozhilov and R.~C.~Rashkov,
  ``Magnon-like dispersion relation from M-theory,''
  arXiv:hep-th/0607116.
  %%CITATION = HEP-TH/0607116;%%

%\cite{Bobev:2006fg}
\bibitem{Bobev:2006fg}
  N.~P.~Bobev and R.~C.~Rashkov,
  ``Multispin giant magnons,''
  Phys.\ Rev.\  D {\bf 74}, 046011 (2006)
  [arXiv:hep-th/0607018].
  %%CITATION = PHRVA,D74,046011;%%
 

%%%%%%% 

%%%%%%%%%%%%%%%%%%%%%%%%%%%%%%

% Hubbard model

%\cite{Rej:2005qt}
\bibitem{Rej:2005qt}
  A.~Rej, D.~Serban and M.~Staudacher,
  ``Planar N = 4 gauge theory and the Hubbard model,''
  JHEP {\bf 0603}, 018 (2006)
  [arXiv:hep-th/0512077].
  %%CITATION = JHEPA,0603,018;%%

%%%%%%%%%%%%%%%%

% anti ferro state

%\cite{Zarembo:2005ur}
\bibitem{Zarembo:2005ur}
  K.~Zarembo,
  ``Antiferromagnetic operators in N = 4 supersymmetric Yang-Mills theory,''
  Phys.\ Lett.\  B {\bf 634}, 552 (2006)
  [arXiv:hep-th/0512079].
  %%CITATION = PHLTA,B634,552;%%

%\cite{Roiban:2006jt}
\bibitem{Roiban:2006jt}
  R.~Roiban, A.~Tirziu and A.~A.~Tseytlin,
  ``Slow-string limit and 'antiferromagnetic' state in AdS/CFT,''
  Phys.\ Rev.\  D {\bf 73}, 066003 (2006)
  [arXiv:hep-th/0601074].
  %%CITATION = PHRVA,D73,066003;%%

%%%%%%%%%%%%%%%%%%%%%%%%%%%%%%%%%%%%%%%%%%%%

\bibitem{flatc}
  C.~J.~Burden,
  ``Gravitational Radiation From A Particular Class Of Cosmic Strings,''
  Phys.\ Lett.\ B {\bf 164}, 277 (1985), \\
  %%CITATION = PHLTA,B164,277;%%
  C.~J.~Burden and L.~J.~Tassie,
  ``Additional Rigidly Rotating Solutions In The String Model Of Hadrons,''
  Austral.\ J.\ Phys.\  {\bf 37}, 1 (1984).
  %%CITATION = AUJPA,37,1;%%

%\cite{Kruczenski:2006pk}
\bibitem{spikyS5}
  M.~Kruczenski, J.~Russo and A.~A.~Tseytlin,
  ``Spiky strings and giant magnons on S**5,''
  JHEP {\bf 0610}, 002 (2006)
  [arXiv:hep-th/0607044].
  %%CITATION = JHEPA,0610,002;%%

%%%%%%%%%% N-R system %%%%%%%%%55

%\cite{Arutyunov:2003za}
\bibitem{AFRT}
G.~Arutyunov, S.~Frolov, J.~Russo and A.~A.~Tseytlin,
``Spinning strings in AdS(5) x S5 and integrable systems,''
Nucl.\ Phys.\ B {\bf 671}, 3 (2003)
[arXiv:hep-th/0307191]
%%CITATION = HEP-TH 0307191;%%


\bibitem{ART}
  G.~Arutyunov, J.~Russo and A.~A.~Tseytlin,
  ``Spinning strings in AdS(5) x S5: New integrable system relations,''
  Phys.\ Rev.\ D {\bf 69}, 086009 (2004)  [arXiv:hep-th/0311004].
  %%CITATION = HEP-TH 0311004;%%

\bibitem{NR}
 C. Neumann, 
 De problemate quodam mechanico, quod ad primam integralium ultraellipticorum classem revocatur,''
Jour. reine Angew. Math. {\bf 56}, 1859 pp. 46-63, \\
 E. Rosochatius,
 ``\"Uber Bewegungen eines Punktes,''
Dissertation at Univ. G\"otingen, Druck von Gebr. Unger, Berlin 1877.

\bibitem{Moser} 
 J. Moser,
``Various aspects of integrable Hamiltonian Systems,''
in ``Dynamical systems'', Progress in Mathematics {\bf 8}, 
C.I.M.E. Lectures, Bressanone, Italy, (1978), J. Coates, S.  Helgason, Eds.
 
%%%%%%%%%%%%%%%%

% More Recent



\bibitem{recent}
  J.~Kluson and K.~L.~Panigrahi,
  ``D1-brane in beta-Deformed Background,''
  arXiv:0710.0148 [hep-th], \\
  %%CITATION = ARXIV:0710.0148;%%
  H.~Dimov and R.~C.~Rashkov,
  ``On the anatomy of multi-spin magnon and single spike string solutions,''
  arXiv:0709.4231 [hep-th], \\
  H.~Hayashi, K.~Okamura, R.~Suzuki and B.~Vicedo,
  ``Large Winding Sector of AdS/CFT,''
  arXiv:0709.4033 [hep-th], \\
  P.~Bozhilov and R.~C.~Rashkov,
  ``On the multi-spin magnon and spike solutions from membranes,''
  arXiv:0708.0325 [hep-th], \\
  %%CITATION = ARXIV:0708.0325;%%
  N.~P.~Bobev and R.~C.~Rashkov,
  ``Spiky Strings, Giant Magnons and beta-deformations,''
  Phys.\ Rev.\  D {\bf 76}, 046008 (2007)
  [arXiv:0706.0442 [hep-th]].
  %%CITATION = PHRVA,D76,046008;%%

%%%%%%%%%%%%%%%%%%%%%%%%%%%%%%%%%%%%%%%%%%%%%%%5

% Simultaneous

%\cite{Mosaffa:2007ty}
\bibitem{Mosaffa:2007ty}
  A.~E.~Mosaffa and B.~Safarzadeh,
  ``Dual Spikes: New Spiky String Solutions,''
  arXiv:0705.3131 [hep-th].
  %%CITATION = ARXIV:0705.3131;%%


%-----------------------------------------------------------------------------------
%%%%%%%%%%%%%%%%%%%%%%%%%%%%%%%%%%%%%%%%%%%%%%%%%%%%%%%%%%%%%%%%%%%%%%%%%%%%%%%%%%%
%%%%%%%%%%%%%%%%%%%%%%%%%%%%%%%%%%%%%%%%%%%%%%%%%%%%%%%%%%%%%%%%%%%%%%%%%%%%%%%%%%%
%%%%%%%%%%%%%%%%%%%%%%%%%%%%%%%%%%%%%%%%%%%%%%%%%%%%%%%%%%%%%%%%%%%%%%%%%%%%%%%%%%%
%%%%%%%%%%%%%%%%%%%%%%%%%%%%%%%%%%%%%%%%%%%%%%%%%%%%%%%%%%%%%%%%%%%%%%%%%%%%%%%%%%%
%-----------------------------------------------------------------------------------



\end{thebibliography}
\end{document}